\documentclass[12pt]{article}
\usepackage{epsfig}
\usepackage{array,tabularx}
\usepackage[english]{babel}

\makeatletter
\def\@biblabel#1{}
\def\@cite#1#2{{#1\if@tempswa , #2\fi}}
\makeatother
\input{tcilatex}

\begin{document}

% \maketitle
%===================   TITLE

\begin{center}
\textbf{\Large Pulsars Born in Open Stellar Clusters}

\vskip 3mm

{\large {Pskovskii Yu. P.$^1$, Dorofeyev O.F.$^2$}}\\[0pt]

\vskip 2mm $^1$Professor of Sternberg Astronomical Institute, Moscow State
University (MSU), Vernadskiy Avenue, 13, Moscow, Russia. Phone No.
(095)~939--16--83.\\[5pt]
$^2$ Assistant Professor of Physics Faculty, Moscow State University (MSU),
Vorobiovy Gory, Moscow, Russia. Phone No. (095)~939--31--77.
\end{center}

%
%
%+++++++++++++++++++++++

\abstract{For two nearby  pulsars, formed close to the galactic plane,
their host open stellar clusters have been found.} \newpage

\section{ Birth site for pulsar B\,2224 + 65.}

Cordes et al. (1993) discovered the complex consisting of pulsar B\,2224 +
65 and of the gaseous nebula ``Guitar'', the pulsar being located at the end
of its ``neck''. The pulsar is specified by a great proper motion which may
be due to its proximity to the Sun or, otherwise, it is a very peculiar
object even among the pulsars. The only procedure used so far to measure the
distance using the dispersion measure of the pulsar, $DM\,=\,35.3$~ps/cm$%
^{-3}$, yields $D_0\sim 1.95$~kps. There is, however, a ring of gas around
the pulsar ionized by a shock wave (the density jump is characterized by a
factor of $10 \div 12$~(see Kaplan \& Pikelner, 1979)). Electrons which
contribute to the dispersion measure of the pulsar may partly belong to this
gaseous ring. Therefore, the value $D_0\,\sim\,1.95$~kps seems to be
overestimated. The discovery of soft X--ray radiation from the nebula
``Guitar'', that contains the moving pulsar (Romani et al., 1997), also
provides evidence for the complex to be close to the Sun. A more precise
distance to the pulsar could be estimated from its trigonometric parallax
when available.

Characteristic pulsar age is $t_c\,=\,0.5\,P/\dot P\,=\,1.12$ Myr (Taylor et
al., 1993) where $P$ is the pulse period, $\dot P$, its first derivative in
seconds. Characteristic ages of pulsars are found to be systematically high.
A real age is determined more precisely using the expression for so--called
``kinematic age'' (Smith, 1977): 
\[
t_k\,=\,0.5\,\tau_D\,ln(1\,+\,t_c/(0.5\,\tau_D)). 
\]
The parameter of magnetic field attenuation is estimated by us (Pskovskii \&
Dorofeyev, 2001) to be $\tau_D\,=\,1.8\,\pm\,0.2$ Myr. Thus, kinematic age
of pulsar 2224+65 is $t_k\,=\,0.74\,\pm\,0.10$ Myr.

The position angle $PA$ of the pulsar motion has been preliminarily found to
be $52^o$ which indicates that the pulsar has arrived at its current site on
the celestial sphere from the IIIrd quadrant, that is, from the area within $%
\alpha(19^h\,\div\,22^h), \, \delta(+70^o\,\div\,65^o)$. To determine the
site of its birth more accurately one needs to take into account systematic
variations in pulsar radial velocity and proper motion, though its spatial
velocity may be regarded as being constant.

If the pulsar has been moving towards the Sun for the past $0.7$ Myr, it is
likely to have started from the region close to an open stellar cluster
localized not far from the Sun in the same quadrant. A possible candidate is
cluster M\,39 (NGC\,7092).

\section{Formulae for calculation of the pulsar starting site}

A trajectory of motion for a pulsar having a constant spatial velocity is
rectilinear, but its projection on the celestial sphere is an arc of a great
circle. We shall use the rectilinear equatorial reference frame $XYZ$ with
the origin $O$ at the point of observations (Fig. 1). The current position $%
\Pi(\alpha_o,\,\delta_o,\,1950)$ of the pulsar belongs to the plane $YZ$. It
corresponds to the position $\Pi(x_o,\,y_o,\,z_o)$ in the rectilinear frame,
where $x_o\,=\,0$, $y_o\,=\,D_o\,\cos\delta_o$, $z_o\,=\,D_o\,\sin\delta_o$.
The components of the pulsar spatial velocity $V$ are directed,
correspondingly \newline

$V_{\alpha}\,(V_{\alpha},\,y_o,\,z_o)$ --- parallel to the axis $OX$, 
\newline

$V_{\delta}(0,\,y_o-V_{\delta}\,\sin\delta_o,\,z_o+V_{\delta}\,\cos\delta_o)$
--- along the meridian, \newline

$V_r(0,\,y_o+V_r\,\cos\delta_o,\,z_o+V_r\,\sin\delta_o)$ --- along the ray $%
\Pi\,O$. \newline

These components are obtained in km/s if expressed in terms of observational
values $V_{\alpha}=4.74\,D_o\,\mu_{\alpha}$, $V_{\delta}=4.74\,D_o\mu_{%
\delta}$, $V_t=4.74\,D_o\,\mu$ (the tangential pulsar velocity for the
current epoch), $V_r=V\,\cos\zeta$. Here $\mu$, $\mu_{\alpha}$, $%
\mu_{\delta} $ are the proper motion and its components along the axes $OX$
and $OY$ expressed in arc seconds per year and corrected for the motion of
the Galaxy and for the motion with respect to the Sun; $\zeta$ is the angle
between the rotational axis of the pulsar and the direction towards an
observer. The expression $V_r=V\,\cos\zeta$ follows from the result
established earlier (Tademaru, 1977; Pskovskii \& Dorofeyev, 1987); the
directions for both $V$ and $V_t$ are supposed to be known from the ``rule
of signs'' for $\zeta$ (Pskovskii \& Dorofeyev, 2001).

The system of equations for a straight line parallel to the vector 
\[
V(V_{\alpha },V_{r}\cos \delta _{o}-V_{\delta }\sin \delta _{o},V_{r}\sin
\delta _{o}+V_{\delta }\cos \delta _{o}) 
\]
is given by 
\begin{equation}
\frac{x}{V_{\alpha }}=\frac{y-y_{o}}{V_{r}\cos \delta _{o}-V_{\delta }\sin
\delta _{o}}=\frac{z-z_{o}}{V_{r}\sin \delta _{o}+V_{\delta }\cos \delta _{o}%
}=0.978\,t_{k}=\vartheta ,  \label{parallel}
\end{equation}
where $x,y,z$ are the rectilinear coordinates of a pulsar at a starting
moment (in parsecs), $0.978=\vartheta /t_{k}$ is a dimensional factor in
units of ps/km/s. By their sense, the values $t_{k}$ and $\vartheta $ are
negative. As the value $D_{o}$ is not known, and neither are the velocity
and its components expressed in terms of $D_{o}$, the system (\ref{parallel}%
) can be transformed as follows: 
\begin{eqnarray}
x^{\prime }/D_{o} &=&4.74\vartheta \mu _{\alpha },  \nonumber
\label{transform} \\
y^{\prime }/D_{o} &=&(1\pm 4.74\vartheta \mu \cot \zeta )\cos \delta
_{o}-4.74\vartheta \mu _{\delta }\sin \delta _{\delta }, \\
z^{\prime }/D_{o} &=&(1\pm 4.74\vartheta \mu \cot \zeta )\sin \delta
_{o}+4.74\vartheta \mu _{\delta }\cos \delta _{\delta },  \nonumber \\
(D^{\prime }/D_{o})^{2} &=&(x^{\prime }/D_{o})^{2}+(y^{\prime
}/D_{o})^{2}+(z^{\prime }/D_{o})^{2}.  \nonumber
\end{eqnarray}
$D^{\prime }$ is the pulsar distance at a starting moment. The equatorial
coordinates for the starting moment $\alpha ^{\prime },\delta ^{\prime }$
are given by:

\begin{eqnarray}
\alpha ^{\prime } &=&\alpha _{0}+\frac{y^{\prime }}{\cos \delta ^{\prime }}%
\arcsin \left( \frac{x^{\prime }}{D^{\prime }}\right) ,  \nonumber \\
\delta ^{\prime } &=&\arctan (\frac{z^{\prime }}{y^{\prime }}).  \nonumber
\end{eqnarray}
The coordinates $x_{M}^{\prime },y_{M}^{\prime },x_{M}^{\prime };\alpha
_{M}^{\prime },\delta _{M}^{\prime }$ for cluster M\thinspace 39 at the
moment $-t_{k}$ are obtained using the formulae (\ref{parallel}).

\bigskip

\section{The birth site of pulsar B\,2224+65}

The parameters of pulsar B\,2224+65 and of the open stellar cluster M\,39
which are currently observed, as well as those calculated for the moment of
birth, are shown in the second and the third columns of Table 1,
respectively. The other columns of the Table contain findings for the
objects to be reviewed below. The nomenclature of the rows is explained in
the text. The data for the pulsars are given according to Taylor et al.
(1993), and those for the open stellar clusters, according to Rastorguev \&
Glushkova (1999). The angle $\vartheta$ has been determined by means of our
approach (Pskovskii \& Dorofeyev, 1998).

Since the distance $D_o$ to pulsar B\,2224+65 is not known, the values of $%
x^{\prime}/D_o$, $y^{\prime}/D_o$, etc. are to be calculated. For the rest
of the pulsars and for all the stellar clusters the values $x^{\prime}$, $%
y^{\prime}$, etc. have been determined as the distances $D_o$ are available
for them.

Because of significant errors involved in the technique of determining
initial coordinates of a pulsar using its current estimated values $\mu$, $%
V_r$, $D_o$, and $\vartheta$, one can anticipate, in the best case, but the
full (or a partial) superposition of an error box of a pulsar and that of a
stellar cluster. In our case, the distance to pulsar B\,2224+65 estimated at
the moment of birth cannot be matched with that for the stellar cluster
M\,39.

\begin{tabular}{ccccc}
Table 1 &  &  &  &  \\ 
&  &  &  &  \\ 
Objects & PSR\ B2224+65 & M39 & PSR B192929+10 & NGC 6633 \\ 
$l_{0},\;b_{0}$ & 108.6,\ \ +6.8 & 92.5,\ \ -2.3 & 47.4,\ \ -3.9 & 36.1,\ \
-8.8 \\ 
$\alpha _{0}\left( h,m\right) $ & 22$^{h}24^{m}$ & 21$^{h}30.4^{m}$ & 19$%
^{h}29.9^{m}$ & 18$^{h}25.5^{m}$ \\ 
$\delta _{0}\left( ^{\circ }\right) $ & +65.34 & +48 .22 & +10.88 & +6.53 \\ 
$\mu _{\alpha }\left( ^{\prime \prime }/yr\right) $ & 0.146$\pm 0.003$ & 
=0.0504$\pm 0.0013$ & 0.096$\pm 0.006$ & 0.000$\pm 0.010$ \\ 
$\mu _{\delta }\left( ^{\prime \prime }/yr\right) $ & 0.113$\pm 0.003$ & 
-0.016$\pm 0.003$ & 0.050$\pm 0.004$ & 0.002$\pm 0.010$ \\ 
$\mu \left( ^{\prime \prime }/yr\right) $ & 0.185$\pm 0.003$ & 0.0191$\pm
0.0013$ & 0.108$\pm 0.006$ & 0.002$\pm 0.010$ \\ 
$\varsigma \left( ^{\circ }\right) $ & 41$\pm 3.0$ & - & 24.4$\pm 3.0$ & -
\\ 
$\phi \left( ^{\circ }\right) $ & - & 1.5 & - & 1.3 \\ 
$D_{0}\left( pc\right) $ & $\left( 200\pm 50\right) $ & 350$\pm 100$ & 170$%
\pm 40$ & 380$\pm `00$ \\ 
$V_{t}\left( km/s\right) $ & $\left( 200\pm 30\right) $ & 32$\pm 7$ & 90$\pm
20$ & 4$\pm 20$ \\ 
$V_{r}\left( km/s\right) $ & $\left( -230\pm 30\right) $ & -7.5$\pm 5$ & -190%
$\pm 50$ & -28$\pm 2$ \\ 
$V\left( km/s\right) $ & $\left( -300\pm 40\right) $ & 33$\pm 7$ & 210$\pm 50
$ & 30$\pm 3$ \\ 
$t_{k}\left( Myr\right) $ & 1.12 & - & 3.09 & - \\ 
$t_{c}\left( Myr\right) $ & 0.74$\pm 0.010$ & - & 1.34$\pm 0.16$ & - \\ 
$\vartheta $ & +0.74$\pm 0.010$ & - & -90$\pm 30$ & - \\ 
$\left( x^{\prime }/D_{0}\right) ,\;x^{\prime }\left( pc\right) $ & $\left(
-0.49\pm 0.07\right) $ & 12$\pm 6$ & 400$\pm 80$ & 0$\pm 20$ \\ 
$\left( y^{\prime }/D_{0}\right) ,\;y^{\prime }\left( pc\right) $ & $\left(
1.13\div 0.07\right) $ & 220$\pm 90$ & 30$\pm 50$ & 410$\pm 100$ \\ 
$\left( z^{\prime }/D_{0}\right) ,\;z^{\prime }\left( pc\right) $ & $\left(
1.41\pm 0.010\right) $ & 280$\pm 110$ & 410$\pm 80$ & 40$\pm 30$ \\ 
$\left( D^{\prime }/D_{0}\right) ,\;D^{\prime }\left( pc\right) $ & $\left(
1.87\pm 0.09\right) $ & 360$\pm 100$ & 3.7$\pm 20$ & 410$\pm 100$ \\ 
$\delta ^{\prime }\left( ^{\circ }\right) $ & 51$\pm 11$ & 53$\pm 33$ & 3.7$%
\pm 2.0$ & 1.4$\div 3.6$ \\ 
$\Delta \alpha ^{\prime }\left( h,m\right) $ & 1$^{h}07^{m}\pm 32^{m}$ & $%
+12^{m}\pm 9^{m}$ & $54^{m}\pm 18^{m}$ & $0\pm 0.1^{m}$ \\ 
$\alpha ^{\prime }\left( h,m\right) $ & 21$^{h}17^{m}\pm 32^{m}$ & 21$%
^{h}45^{m}\pm 9^{m}$ & 18$^{h}36^{m}\pm 18^{m}$ & 18$^{h}25.5^{m}\pm 0.1^{m}$%
\end{tabular}

Therefore, the knowledge of possible uncertainties in $\alpha^{\prime} $ and 
$\delta^{\prime}$ is of great importance. To estimate them, the value of $%
D^{\prime}/D_o$ was suggested to possess a relative error which corresponds
to the ``bb'' class in the classification proposed by Taylor et al. (1993),
that is, 46 per cent. Errors of the values in the formulae have been
suggested to be mean--square ones deduced from expressions for corresponding
functions.

The error box for the stellar cluster M\,39 has been calculated taking into
account the angular diameter of the cluster having an extended corona with a
diameter of $\sim 1.5^o$ (Artiukhina, 1970; Barkhatova \& Pylskaya, 1978).

Fig. 2 shows the error boxes of these objects to be partly superimposed at
the starting moment of the pulsar. The pulsar therefore can be considered to
be at the distance $D^{\prime}$ at the moment in question, this distance
corresponding to cluster M\,39. This fact allows us to calculate the values $%
D_o$, $V$, $V_t$, and $V_r$ for the pulsar. The findings are shown in
Table~1 in brackets. The current distance to the pulsar, $D_o=200\pm 50$~ps,
is obtained to be dramatically small, whereas kinematic parameters are shown
to have rather common values. A dispersion measure of the pulsar in such a
case is mainly due to electrons in a shock wave rather than to interstellar
medium.

A spatial trace of the open stellar cluster M\,39 (Fig. 2) is far shorter
than that of pulsar B\,2224+65. The distance to M\,39 has remained
practically the same, but the directions of pulsar and cluster motions are
almost opposite.

Taking into account that M\thinspace 39 is presumably a member of a
kinematic group in Ursa Major (Lloyd Evans \& Meadows, 1964; Eggen, 1965),
an error box of the pulsar starting position proves to lie completely inside
a field of the group. Unfortunately, it is impossible to determine
kinematics of the object before the start, hence we cannot establish with
certainty whether it belongs to the group or not.

\bigskip

\section{The birth site of nearby pulsar1929+10}

The coincidence of the birth site of pulsar B\,2224+65 with the field of the
stellar cluster M\,39, however, does not exclude the possibility that some
single pulsars may be produced outside stellar clusters. A number of young
single pulsars, namely B\,0531+21, B\,0833-45, B\,0611+22, B\,0656+14, etc.,
which have not travelled very far from their starting points are not related
to any stellar clusters. Some pulsars are found to start at high galactic $z$%
--coordinates (Pskovskii \& Dorofeyev, 2001). From statistical findings
correlations were shown to exist among young pulsars and stellar
associations (Mdzinarishvili, 1997). In addition, pulsars discovered in
globular stellar clusters belong to a quite specific group, as well as
millisecond pulsars and those in binary systems.

Let us consider other pulsars which are close to us at the present time and
which have well established estimates of proper motion, the distances $D_o$,
and ages within the limits $0.7<t_k<1.5$~Myr. In addition to pulsar
B\,2224+65 considered above, these conditions are met for the two more
pulsars, B\,1929+10 and B\,1133+16, the latter having unprecedented proper
motion. Pulsar B\,1133+16 is believed to have been born at a high galactic
latitude, both for $V_r>0$ and for $V_r<0$, so it has no host stellar
cluster.

Unlike pulsar B\,2224+65, determinations of the birth sites of the two
pulsars mentioned above are based on distances established rather well by
Taylor et al. (1993). The positions of these pulsars at a starting moment
have been calculated using formulae (1).

The data concerning a birth site of the pulsar and the parameters of its
host cluster NGC\thinspace 6638 are given in the corresponding columns of
Table 1. The value of $\zeta $ has been estimated by our technique
(Pskovskii \& Dorofeyev, 1998). Cluster NGC\thinspace 6633 has low proper
motion, its error box lying almost entirely within that of pulsar
B\thinspace 1929+10 at the moment of its start (Fig. 3).

\bigskip 

\section{Conclusion}

It has been shown that two nearby pulsars, which started from the galactic
plane, may have been related, at a moment of their birth, to open stellar
clusters. It is worthwhile pointing out that the both pulsars were reported
to be weak sources of X--rays (Wang et al., 1993; Cordes et al., 1993).

There is good reason to believe that actually there are more pulsars related
to open clusters. This can be deduced from the work of Mdzinarishvili (1997)
containing evidence for the correlation of young pulsar ($t_c<2$ Myr)
distribution close to the galactic plane with that of O--associations in the
vicinity of the Sun.

The authors wish to express gratitude to the researches of the Astronomical
Department of the MSU Physics Faculty, Professor A. S. Rastorguev and
Assistant Professor Ye. V. Glushkova, who have given the data for the open
stellar clusters.

\newpage

\centerline{\underline{Figure captions}} \bigskip 
% to the article by Yu. P. Pskovskii and O. F. Dorofeyev \\

Fig.1. Projections of a pulsar radial velocity $V_r$ and of the components $%
V_{\alpha}$ and $V_{\delta}$ of a pulsar tangential velocity $V_t$ on axes
in the rectilinear equatorial reference frame $X,Y,Z$. The case of $V_r>0$
is assumed. The following designations are used: $\Delta
y_1=V_{\delta}\cos\delta_0$, $\Delta z_1=V_{\delta}\sin\delta_0$, $\Delta
y_2=V_r\cos\delta_0$, $\Delta z_2=V_r\sin\delta_0$. \newline

Fig. 2. Positions of pulsar B\,2224+65 (empty squares) and of the open
stellar cluster M\,39 (empty circles) in the sky at the present time. For
the pulsar starting moment the objects are indicated correspondingly by
filled squares and filled circles which lie in the centres of their error
boxes. The dashed line stands for the error box of the cluster; the
continuous line, for the pulsar. The heavy line signifies the galactic
equator. Coordinates are given for the epoch 1950.0. \newline

Fig. 3. Positions of pulsar B\,1929+10 and of the open stellar cluster
NGC\,6633 in the sky. The designations are the same as to Fig. 2.

\end{document}